\documentclass[%
 reprint,
nofootinbib,
 amsmath,amssymb,
 aps,
 onecolumn,
]{revtex4-2}

\usepackage{graphicx}% Include figure files
\usepackage{dcolumn}% Align table columns on decimal point
\usepackage{bm}% bold math
\usepackage{graphicx}
\usepackage{placeins}
\usepackage[pdftex, pdftitle={Article}, pdfauthor={Author},colorlinks = true,allcolors = blue]{hyperref}
\newcommand{\opencharm}{$\gamma p \to \bar{D}^{0}\Lambda_{c}$}
\newcommand{\opencharmprime}{$\gamma p \to \bar{D}^{*}\Lambda_{c}$}

\newcommand{\jpsi}{$J/\psi$}

\begin{document}

\preprint{APS/123-QED}

\title{Letter of Intent: Open Charm at JLab with the sPHENIX MAPS tracker}% Force line breaks with \\

\author{Miguel Arratia }
 \email{Corresponding author: miguel.arratia@ucr.edu}
 \affiliation{University of California, Riverside}%
\author{Sebouh J. Paul}%
 \affiliation{University of California, Riverside}%
\author{Yuri Gotra }
 \affiliation{Thomas Jefferson National Accelerator Facility}%
\author{Hayk Hakobyan }
 \affiliation{Universidad Tecnica Federico Santa Maria, Valparaiso, Chile}%
 \author{Bryan McKinnon }
 \affiliation{University of Glasgow, Glasgow, United Kingdom}%

\date{\today}% It is always \today, today,

\begin{abstract}
We propose a physics program at JLab with CLAS12 focusing on open-charm measurements, aiming to complement and expand current studies of $J/\psi$ at (sub) threshold. This program will aid us in elucidating the $J/\psi$ production mechanisms, which is crucial for interpreting data in terms of gluon form factors and offer potential insights into the intrinsic charm hypothesis and cold-nuclear matter effects. We discuss the technical feasibility of integrating the sPHENIX monolithic-active-pixel sensor (MAPS) tracker, known as MVTX, with the CLAS12 detector. The sPHENIX MTVX would support an open-charm program by providing excellent secondary-vertex performance for tagging $D$ mesons. We study the kinematics of $\gamma p \to \bar{D}^{0}\Lambda_{c}$ through phase-space simulations and estimate rates for the tagged quasi-photoproduction regime available with the CLAS12 forward tagger. While open-charm cross-sections at threshold remain uncertain, various predictions suggest that these measurements could be feasible when combined with conservative estimates of detector acceptance and luminosity. These preliminary estimates motivate detailed Geant detector simulations of signals and backgrounds, along with thorough technical assessments of operating conditions, to further explore the feasibility of these measurements in future dedicated CLAS12 experiments at JLab.

\end{abstract}

\maketitle
\vspace{-0.3cm}
\tableofcontents
\newpage
\section{Motivation for a Measurement of Open-Charm production at Threshold}
\FloatBarrier
The study of \jpsi~photoproduction at threshold has attracted considerable attention, with the objective of accessing the gluonic structure of the nucleon, and, ultimately, contributing to our understanding of the origin of the nucleon mass, see e.g Refs.~\cite{Kharzeev:1998bz,Brodsky:2000zc,Mamo:2019mka,Guo:2021ibg,Kharzeev:2021qkd,Mamo:2022eui,Wang:2022ndz,Guo:2023pqw} . This has motivated various experiments at JLab~\cite{Dudek:2012vr,Joosten:2018gyo}, including GlueX at Hall-D~\cite{GlueX:2019mkq,GlueX:2023pev}, the \jpsi--007 experiment at Hall-C~\cite{Duran:2022xag}, CLAS12 at Hall-B~\cite{Newton:2021whz,Tyson:2023yer}, and SoLID at Hall-A~\cite{JeffersonLabSoLID:2022iod}.

\begin{figure}[h!]
    \centering
        \includegraphics[width=0.3\linewidth]{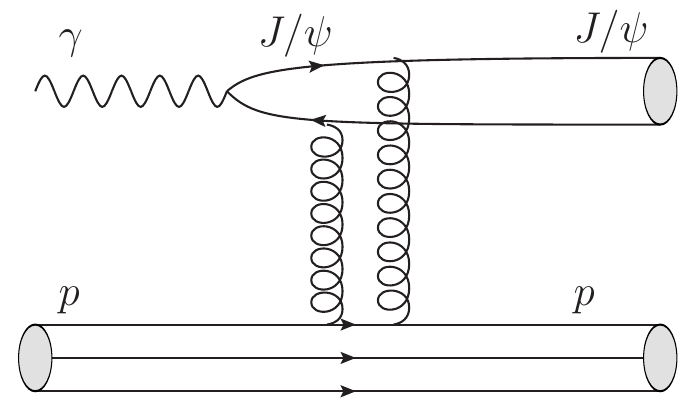}
    \includegraphics[width=0.3\linewidth]{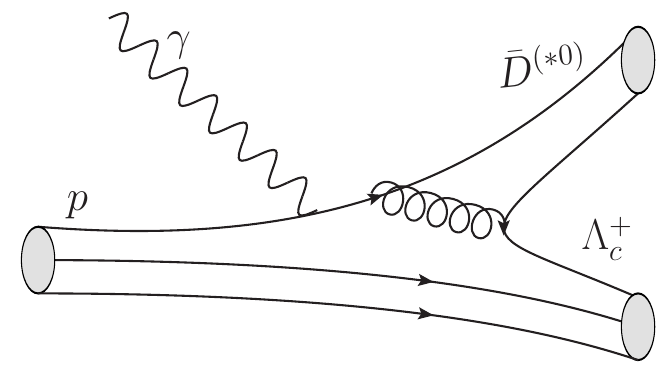}
    \includegraphics[width=0.25\linewidth]{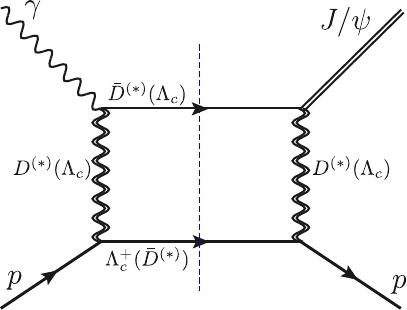}
    \caption{Left: $J/\psi$ production mechanism as per the vector-dominance model. Middle: Open-charm production mechanism. Right: $J/\psi$ production mechanism through open-charm coupled channel suggested by Du \textit{et al}.~\cite{Du:2020bqj}. Source: Ref.~\cite{Du:2020bqj}.}
    \label{fig:diagrams}
\end{figure}

The analysis of $\gamma p \to J/\psi p$ cross-section measurements typically relies on the assumption that the reaction mechanism can be described by the ``vector-dominance model'' (VDM) as illustrated in the left panel of Figure~\ref{fig:diagrams}. Within this framework, the reaction is factorized into a hard $\gamma \to c\bar{c}$ vertex and a soft proton matrix element. Crucially, this factorization enables the extraction of the forward elastic-scattering amplitude of \jpsi~off a proton, see e.g Ref.~\cite{Gryniuk:2016mpk,Strakovsky:2019bev,Pentchev:2020kao,Wang:2022xpw}. 

Such assumptions have, however, been questioned, for example by Du \textit{et al.}~\cite{Du:2020bqj} who suggested that a the reaction mechanism leading to the \jpsi $p$ final state could at least in part proceed through an open-charm coupled channel, illustrated in the right panel of Figure~\ref{fig:diagrams}. In this model, open-charm production is followed by re-scattering of $\bar{D}$ and $\Lambda_{c}$ to produce the the \jpsi $p$ final state. Note that the energy threshold for \jpsi~photoproduction ($\approx$8.2 GeV) is close to the threshold of the \opencharm~and~\opencharmprime~reactions ($\approx$8.7 GeV and $\approx$9.4 GeV, respectively). This mechanism is not factorized like VDM and is not amenable to the interpretation of the $\gamma p \to J/\psi p$ data in relation to the gluonic structure of the nucleon.

\begin{figure}[h!]
    \centering
    \includegraphics[width=0.49\linewidth]{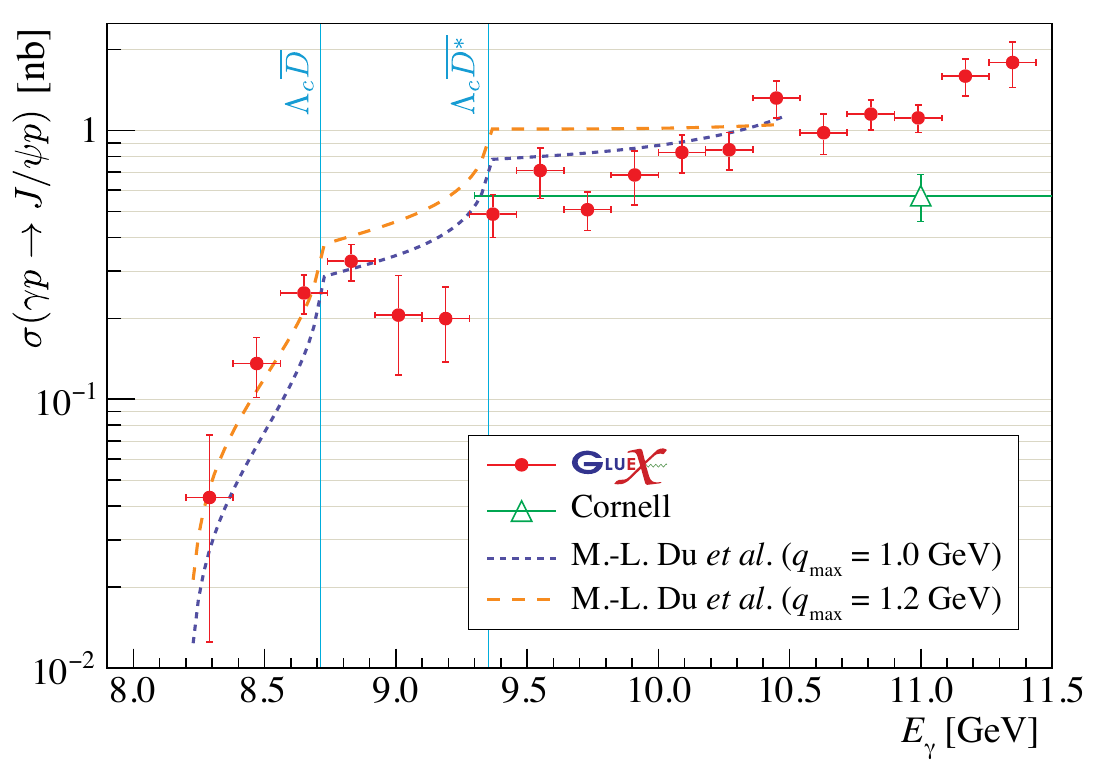}
    \includegraphics[width=0.49\linewidth]{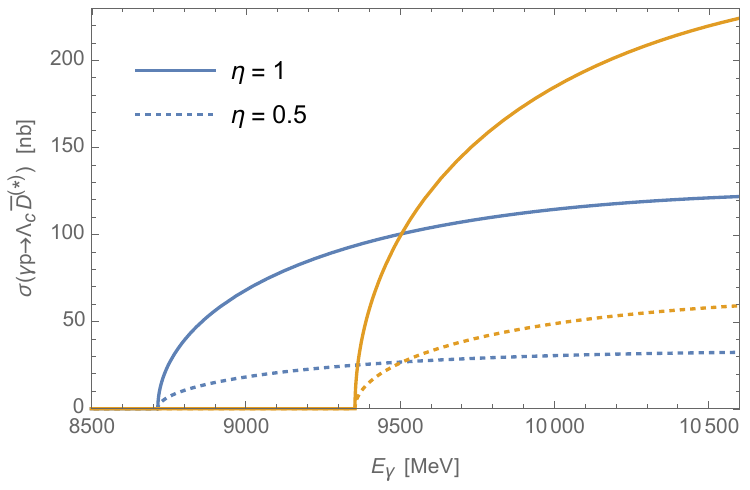}
    \caption{Left: GlueX measured \jpsi~photoproduction near the threshold, with vertical lines indicating the threshold energy for production of \opencharm~and \opencharmprime. The curves represent predictions by Du \textit{et al}.~\cite{Du:2020bqj}. Source: Ref.~\cite{GlueX:2023pev}. Right: Du \textit{et al.} predictions for open-charm production near the threshold, for $\Lambda_c^+\bar D^0$ (blue) and $\Lambda_c^+\bar D^{*0}$ (orange), indicating two scenarios depending on the model parameter $\eta$.  Note how the open-charm cross-section is predicted to be one or two orders of magnitude higher than the \jpsi~cross-section. Source: Ref.~\cite{Du:2020bqj}.}
    \label{fig:GlueXandprediction}
\end{figure}

The latest measurement of \jpsi~at threshold by GlueX~\cite{GlueX:2023pev}, shown in the left panel of Figure~\ref{fig:GlueXandprediction}, revealed intriguing trends in the dataset near threshold, although with weak statistical significance (2.6$\sigma$). These features, not initially anticipated within the vector-dominance model, align with the open-charm mechanism proposed by Du \textit{et al.}\cite{Du:2020bqj}. Notably, the cusp features appear at the threshold values for both \opencharm~and~\opencharmprime. 
\FloatBarrier
Another intriguing feature not expected in the VDM model was uncovered by the Hall-C experiment~\cite{Duran:2022xag}, which measured differential double-differential cross-sections as a function of the momentum transfer, $t$, and photon energy, $E_{\gamma}$. For the most part, the cross-section is well described by a falling exponential function in $t$, but some data present an upward trend at high $t$, which might be expected from $u$-channel exchange. 

A partial-wave analysis carried out by the JPAC collaboration~\cite{JointPhysicsAnalysisCenter:2023qgg} suggests that both features—namely, the cusps in GlueX data and the upward trend in $t$ in Hall-C data—are compatible with the open-charm coupled channel model, although with large statistical uncertainties. Thus far, the Du \textit{et al.} model cannot be ruled out with existing experimental data, or as they put it: {\it ``Our analysis indicates that the present statistics do not exclude severe violations of factorization and of the Vector Meson Dominance which are usually assumed in the literature''}.

One way to indirectly constrain the open-charm channel scenario would be through future, higher-statistics \jpsi~measurements at GlueX, Hall-C, and SoLID. A direct alternative arises from the observation that the strength of the cusp structures and overall cross-sections in GlueX \jpsi~data is related to the open-charm cross-sections., \opencharm, which Du \textit{et al}. predict to range between one and two orders of magnitude higher than that of $\gamma p\to$\jpsi $p$, depending on model parameters, as depicted in the right panel of Figure~\ref{fig:GlueXandprediction}. Therefore, one way to constrain the extent to which the open-charm exchange mechanism contributes to the total \jpsi~cross-section is by measuring the open-charm cross-section itself, i.e \opencharm~and~\opencharmprime. No previous measurement for open-charm photoproduction exists for photon energies below 20 GeV, making experiments at JLab 12 GeV rich in discovery potential. 

The importance of directly measuring open-charm photoproduction near threshold has been emphasized in recent work, as evidenced by the following sample of quotes:

\begin{quote}
    \it{``Since the strength of the cusps is connected to the rate for \opencharm, we also provide an estimate for the expected rate into the open-charm channels....measurements of the $\bar{D}\Lambda_{c}$ production will provide crucial information}''. Du \textit{et al}.~\cite{Du:2020bqj}.
\end{quote}

\begin{quote}
\it{``It is thus crucially important to constrain model parameters with further measurements in order to disentangle the possible physics scenarios and their implications...the measurement of open-charm photoproduction is needed to assess the role of coupled channels. A simultaneous analysis of the $\gamma p\to$\jpsi~and \opencharm~cross sections would provide a stringent constraint on the coupled channel dynamics. Based on the best fit parameters extracted here, we expect a large open-charm cross-section $\gtrsim$ 10 nb. Furthermore, studies of photoproduction off nuclear targets may give further constrain on the total \jpsi-nucleon cross-section.'' } Winney \textit{et al}. (JPAC Collaboration)~\cite{JointPhysicsAnalysisCenter:2023qgg}.
\end{quote}

The cross-section for open-charm photoproduction near threshold is currently unknown, but measurements have been conducted at higher energies. The lowest-energy studies were performed at SLAC, where an experiment using a 20 GeV real photon reported a total open-charm cross-section of $56^{-23}_{+24}$ nb~\cite{SLACHybridFacilityPhoton:1983yfx, Abe:1983pe}. Another SLAC study used a 10.5 GeV photon beam and reported an upper limit of 94 nb at a 90\% confidence level~\cite{SLACHybridFacilityPhoton:1984qik}. The dominant channel for the energy range of interest is \opencharm, which is expected to dominate the total charm cross-section. The HERMES experiment also conducted searches for open-charm production in low-energy photoproduction, with unpublished measurements~\cite{Volk:2001kd} of $D^{*\pm}$ production suggesting a total open-charm cross-section of:
\begin{equation}
    \left[87.9^{+40.7}_{-32.1} \mathrm{(stat)}\pm9.2 \mathrm{(sys)} \pm 17.6 \mathrm{(frag.~model)}\right] \mathrm{nb}~\mathrm{at}~E_{\gamma} = 15~\mathrm{GeV}
\end{equation} 

Tomasi-Gustafsson~\cite{Tomasi-Gustafsson:2004hel} predicts the \opencharm~cross-section to be approximately 40 nb at a photon energy of $10$ GeV. As shown in Fig.~\ref{fig:GlueXandprediction}, Du \textit{et al.} predict an open-charm cross-section within the range of 50--100 nb for the 9--10 GeV range under one assumption, and 20--30 nb under an alternative set of model parameters. For comparison, the GlueX data reveals a $\gamma p \rightarrow J/\psi$ cross-section in the 9--10 GeV range that is approximately 0.3--1 nb (as shown in Figure~\ref{fig:GlueXandprediction}).
\section{Experimental Setup}

We envision taking advantage of the lifetime of the charmed hadrons by using a cut in the displaced vertex position relative to a thin target or prompt particles.  The charmed mesons and baryons will travel $\mathcal O(100\,\mu{\rm m})$ before decaying.  To measure this requires two key ingredients in our proposed experimental setup.  The first is a thin target with a thickness not much greater than 100$\mu{\rm m}$. The second requirement is we need to have a vertex resolution that is much smaller than the decay lengths.  In the current CLAS12 setup, the vertex resolution is insufficient for this purpose.  Instead, we will propose to borrow an existing detector, the MAPS Vertex tracker (MVTX) from the sPHENIX experiment, and insert it inside the CLAS12 Central Vertex Tracker. This displaced-vertex strategy is much more promising than relying solely on the CLAS12 PID systems, which would serve as the baseline for these searches but have limited momentum range and coverage. This possible detector setup is described in detail Sec.~\ref{sec:MVTX} below, and we describe its integration in CLAS12 in Sec.~\ref{sec:integration}.

\subsection{The sPHENIX MAPS Vertex Tracker}
\label{sec:MVTX}
sPHENIX is an ongoing experiment at RHIC~\cite{PHENIX:2015siv}, primarily dedicated to the study of quark-gluon plasma. At the heart of sPHENIX lies the Monolithic-Active-Pixel-Sensor (MAPS) vertex detector known as MVTX~\cite{MVTX}. This detector is designed to track the vertices of particles generated from proton-proton and nucleus-nucleus collisions. More specifically, MVTX plays a crucial role in a physics program centered on the measurement of displaced vertices as signatures of heavy-quarks (charm and bottom) mesons and baryons. 
For reference, the single-track performance of MVTX within the 1.4~T solenoid field of sPHENIX is 40~$\mu$m for pointing resolution of distance-of-closest approach (DCA) for tracks with 0.5--1 GeV transverse momentum, and an efficiency of 80$\%$.

\begin{figure}[h!]
    \centering
    \includegraphics[width=0.35\textwidth]{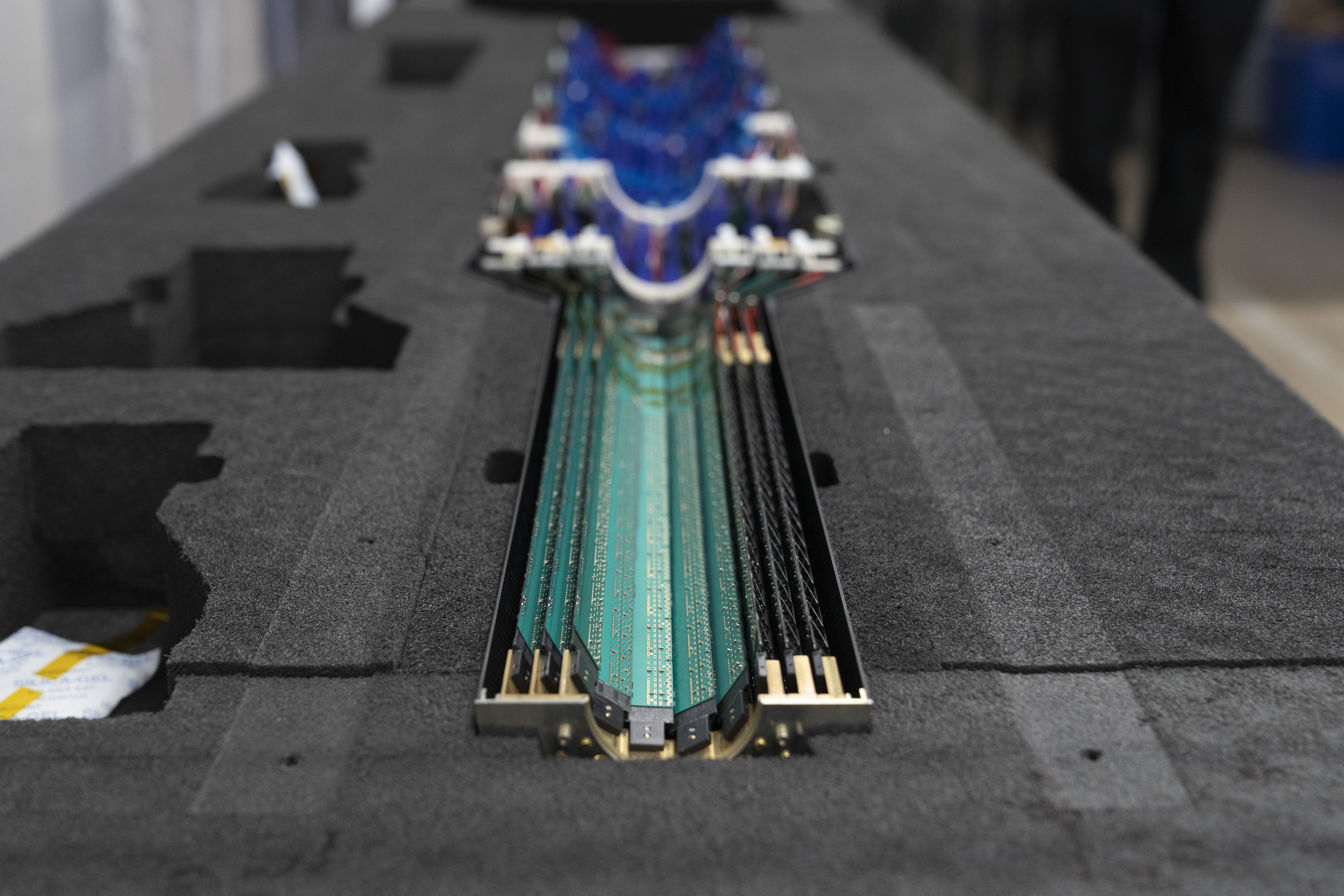}
    \caption{Half of the sPHENIX MVTX detector and services prior to installation in sPHENIX. Figure credit: BNL/sPHENIX}
    \label{fig:MVTX}
\end{figure}

The nominal layout of the MVTX in sPHENIX is defined with a barrel geometry, featuring an active length of 27 cm, with layer radial positions ranging from 22.4 to 42.1 mm. Its services are routed through a carbon-fiber cone structure from just one side of the detector. Figure~\ref{fig:MVTX} shows a picture of part of MVTX and services prior to installation in sPHENIX. 

\subsection{Integration of the MVTX in CLAS12}
\label{sec:integration}

\begin{figure}[h!]
    \centering
        \includegraphics[width=0.45\linewidth]{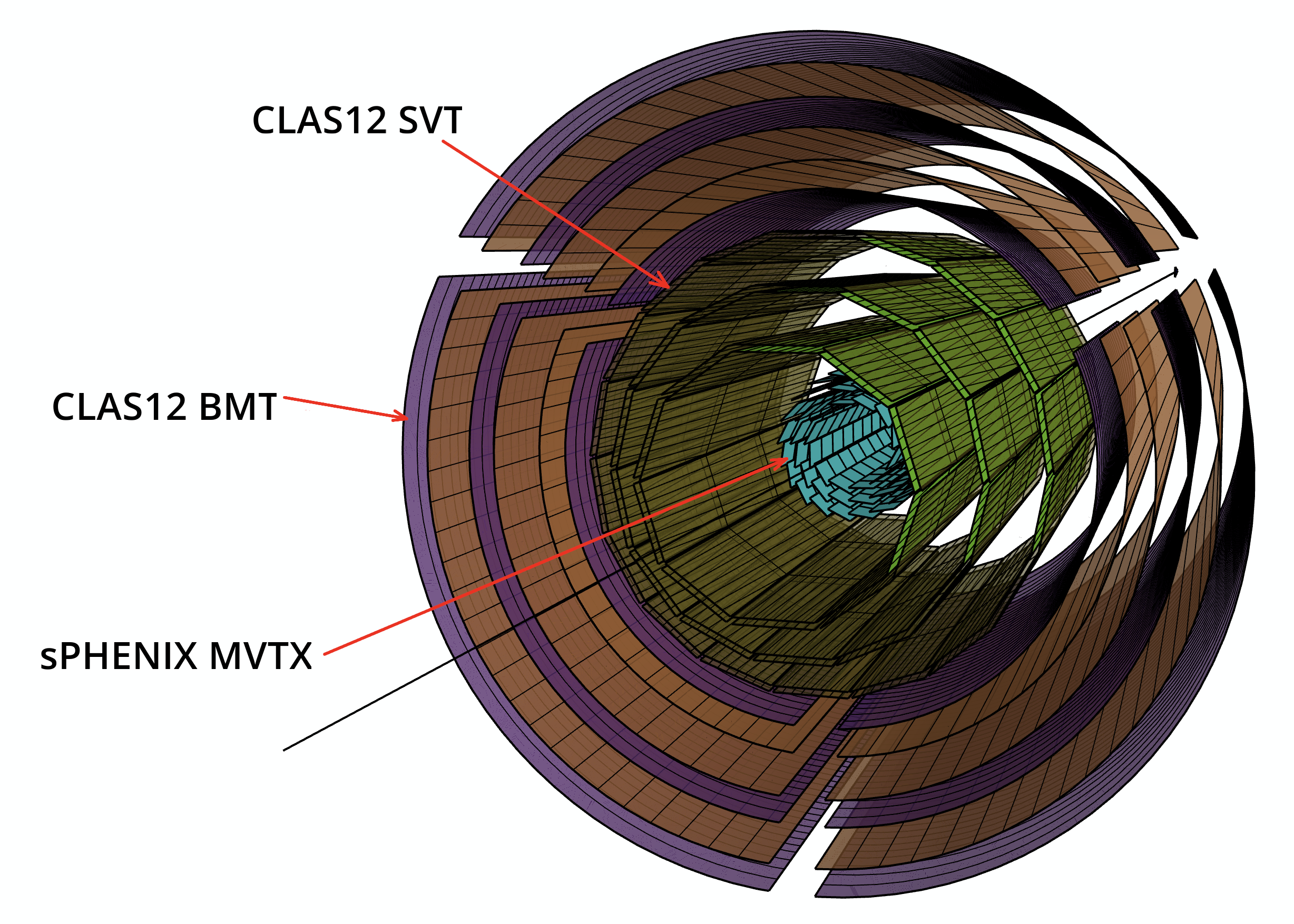}
    \includegraphics[width=0.8\linewidth]{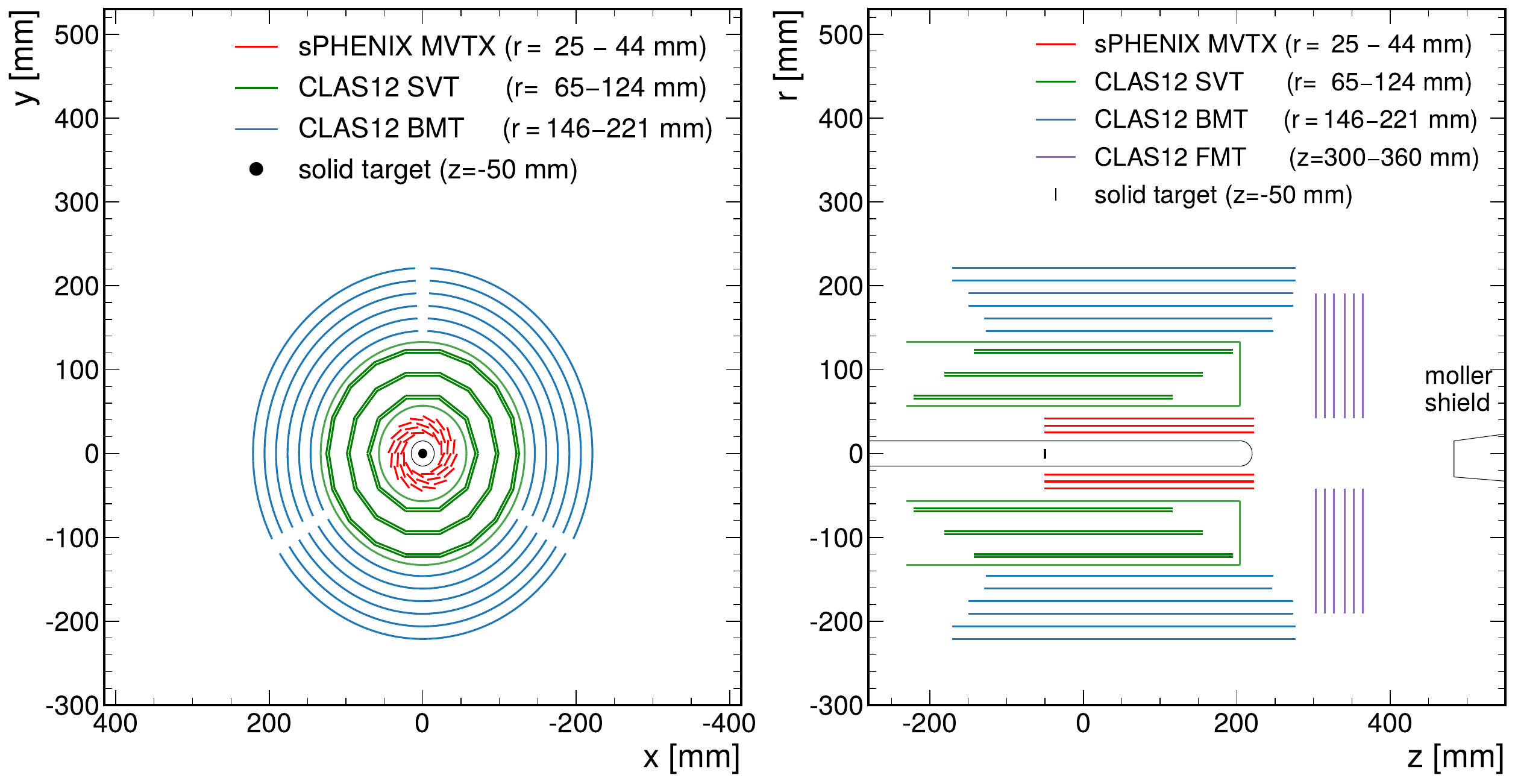}
    \caption{Top: Rendering of possible layout of sPHENIX MVTX (cyan) with the CLAS12 SVT (green) and BMT (purple and orange). Bottom: Possible layout of the sPHENIX MVTX and the CLAS12 SVT and BMT. This configuration would be compatible with a solid-target.}
    \label{fig:sketchup3d}
\end{figure}

We illustrate in Fig.~\ref{fig:sketchup3d} a potential layout for the MVTX within CLAS12, which would use the nominal geometry of the CLAS central trackers (CVT and FMT) and 5 T solenoid, as well as the nominal geometry of MVTX as used in sPHENIX. This layout would be possible by removing the vacuum chamber typically housing a cryotarget for liquid hydrogen or deuterium, thereby creating room for the MVTX detector and its associated services (which would be routed from only one side, as is done in sPHENIX). Instead of having a cryotarget, we would use a smaller vacuum chamber that could host a small solid target. The solid target system could be similar to those used in the CLAS12 run-group D or E experiments, which include several nuclear types in movable setup. The lightest solid target could be beryllium, while the heaviest could be lead or uranium oxide. For reference, the CLAS12 Silicon Verterx Tracker (SVT) inner-most layer is located at $r$=65 mm, and is enclosed by a Faraday cage.  If we place the MVTX such that its upstream end is at the same $z$ position as the solid target, as shown in Fig.~\ref{fig:sketchup3d}, then the polar-angle acceptance of the MVTX is $8.7^\circ<\theta<90^\circ$.  

\subsection{Trigger rate, radiation tolerance, material budget}
The sPHENIX MVTX operates at a trigger rate of 15 kHz at RHIC, which is comparable to the CLAS12 DAQ rate. Additionally, the MAPS sensors have undergone radiation testing up to 2 Mrad, exceeding the anticipated doses from the sPHENIX program, and no radiation damage is expected~\cite{MVTX}, neither during CLAS12 operation. On the other hand, the hit occupancy at CLAS12 would be much smaller compared to RHIC running. Given the MVTX's material budget is less than $<0.5\% X_{0}$ per layer, we expect no detrimental impact on the CLAS12 CVT performance.
\subsection{Triggering with the Forward Tracker}
Possible triggers could be implemented using the forward tagger (FT), which was designed to measure small-angle scattering to access the the small-$Q^{2}$ regime of electro-production ($0.01<Q^{2}<0.1$).  Nominally, the kinematic range of the FT is $2.5^\circ<\theta<4.5^\circ$ for the polar angle~\cite{Acker:2020brf} and it has an energy range of 0.5 to 4.5 GeV.  Therefore, the virtual-photon energy range in this proposed experiment is bound from above by the beam energy minus the minimum energy detectable for the scattered electron, and from below by the $\Lambda_c^+\bar D^0$ threshold.  That is, $8.7<\nu<10.1$ GeV.  
\section{Observables and Analysis Methods}
The physics observable in this analysis will be the cross section of the $\gamma^*p\rightarrow\Lambda_c^+\bar D^{(*)0}$ reaction.  We will not distinguish in this analysis between events in which the $\bar D^0$ is directly produced and those in which it is from the decay of a $\bar D^{*0}$.  If the measured cross section is large enough, we will divide our data sample into bins in the virtual photon energy, $\nu$.  

To measure these cross sections, we will select events with a scattered electron in the FT and candidate of the following charmed hadron decay modes:
\begin{equation}
    \Lambda_c^+\rightarrow K^-\pi^+p\,\,{(\rm B.R.=6.28\pm0.32\%)},
\end{equation}
and 
\begin{equation}
    \bar D^0\rightarrow K^+\pi^-\,\,{(\rm B.R.=3.88\pm0.05\%)}.
\end{equation}
We choose these decays modes because they have a small number of total particles in the final state, all of which are charged, and have relatively large branching ratios.

To reduce background, we will require a cut on the $z$ position of the decay vertex of the charmed-hadron candidate, $v_z$, as obtained using the MVTX\footnote{As an alternative, if there are also prompt particles measured in this reaction, it may be possible to cut on the difference in $v_z$ between the prompt particles and the displaced vertex.}.  We will further reduce background by using a ``bump-hunt'' technique on the invariant-mass distribution of the remaining candidates.  That is, we will fit this distribution to a polynomial (representing any background that remains after the displaced-vertex cut) plus a Gaussian (representing the signal from the charmed particle decay).  

The cross section will then be determined as
\begin{equation}
\sigma_{\gamma p\rightarrow\Lambda_c^+\bar D^{(*)0}}=\frac{N_{\rm fit}}{\ell f_{ep/\gamma p}\epsilon}
\end{equation}
where $N_{\rm fit}$ is the signal yield from the polynomial+Gaussian fit, $\ell$ is the luminosity, $f_{ep/\gamma p}$ is the conversion factor from quasi-real photoproduction\footnote{That is, 
 $ep$ scattering at low $Q^2$} to real photoproduction, and $\epsilon$ is the overall detection efficiency (which includes acceptance effects, detector efficiency, and event-selection cuts) which will be determined using Monte-Carlo simulations.  
\section{Estimated Reach}
In this section, we provide a rough estimation of the experimental reach for the proposed measurement.  These estimates will need to be further refined by future studies using Geant simulations and existing CLAS12 data.
\subsection{Total expected cross section}
Du \textit{et al.} predicted the cross sections for the $\Lambda_c^+\bar D^0$ and $\Lambda_c^+\bar D^{*0}$ and presented the results in Ref.~\cite{Du:2020bqj} with two different values of the model parameter $\eta$.  These two estimates of the cross section differ by a factor of $\approx$4.  We consider the lower estimate of the cross section (corresponding to $\eta=0.5$) to be more realistic, because it is consistent with the experimental upper limit of 94 nb at a 90\% confidence level for total charm production determined by SLAC with a 10.5 GeV photon beam~\cite{SLACHybridFacilityPhoton:1984qik}.  For comparison, the sum of the cross sections for $\bar D^0$ and $\bar D^{*0}$ channels in Du. \textit{et al.} is about 90~nb.  We will therefore base our estimates here on the $\eta=0.5$ calculations.  

The cross section for quasi-photoproduction (that is, electro-production with $Q^2\approx0$) is related to that of real photoproduction by
\begin{equation}
    \sigma_{ep}=\int^{y_{\rm max}}_{y_{\rm min}}\frac{\alpha}{\pi y}\sigma_{\gamma p}(\nu)\left((1 - y + 1/2 y^2) \log\left(\frac{Q^2_{\rm max}}{Q^2_{\rm min}}\right) + m_e^2y^2(Q^{-2}_{\rm max}-Q^{-2}_{\rm min})\right)
    \label{eq:epformula}
\end{equation}
where $y=\nu/E_{\rm beam}$, $y_{\rm min}$ is evaluated at the threshold value of $\nu$ and $y_{\rm max}$ is determined by the minimum detectable scattered-electron energy.  The limits on $Q^2$ for a given $y$ are given by $4 E_{\rm beam}^2(1-y)\sin{\theta^e_{\rm min}}$ and $4 E_{\rm beam}^2(1-y)\sin{\theta^e_{\rm max}}$, where the limits on $\theta_e$ are determined by the acceptance of the FT (2.5$^\circ$ to 4.5$^\circ$).  

This evaluates to $\approx$0.0046 nb with $\bar D^0$ and $\approx$0.0036 nb with $\bar D^{*0}$, which combined is $\approx$0.0082 nb.

Since similar reactions on neutrons (\textit{e.g} $\gamma n\rightarrow \Lambda_c^+D^-$) are expected to be at least an order of magnitude less than those on protons~\cite{Tomasi-Gustafsson:2004hel}, the per-nucleon cross section would be about $Z/N$ times smaller than the free-proton cross-section.  This evaluates to a factor of 
%1/2 for $^{12}$C. 
4/9$\approx$0.44 for $^9$Be.  

\subsection{Luminosity}
To provide ample luminosity while simultaneously preventing the target foil from being so thick that charmed hadrons would decay before exiting the target, we could use a multi-foil target, as was done in Run-Group M with a multi-foil carbon target.  Using a 5-foil Be target with 100~$\mu$m for each foil and an 85 nA beam would correspond to 3.0$\times$10$^{34}$~cm$^{-2}$s$^{-1}$.  For reference, in the currently-running Run-Group E experiment at CLAS12, the luminosity for solid-target production runs are 1.0$\times$10$^{35}$~cm$^{-2}$s$^{-1}$ for carbon (1.5~mm thick foil at 85~nA) and 4.3$\times$10$^{34}$~cm$^{-2}$s$^{-1}$ for lead (143~$\mu$m thick target with 70~nA).

 100 days of running at $3.0\times10^{35}$~cm$^{-2}$s$^{-1}$ and PAC efficiency of 50\% corresponds to 1.62$\times 10^8$~nb$^{-1}$ of luminosity.  Multiplying this by the cross sections in the previous section (and including a factor of 0.44 for the proton fraction) yields $\approx$270k $\Lambda_c^+\bar D^0$ events and $\approx$210k $\Lambda_c^+\bar D^{*0}$ events, which combined is $\approx$470k events. 
 
 \subsection{Monte-Carlo Simulations}
\begin{figure}
     \centering
     \includegraphics[width=0.5\textwidth]{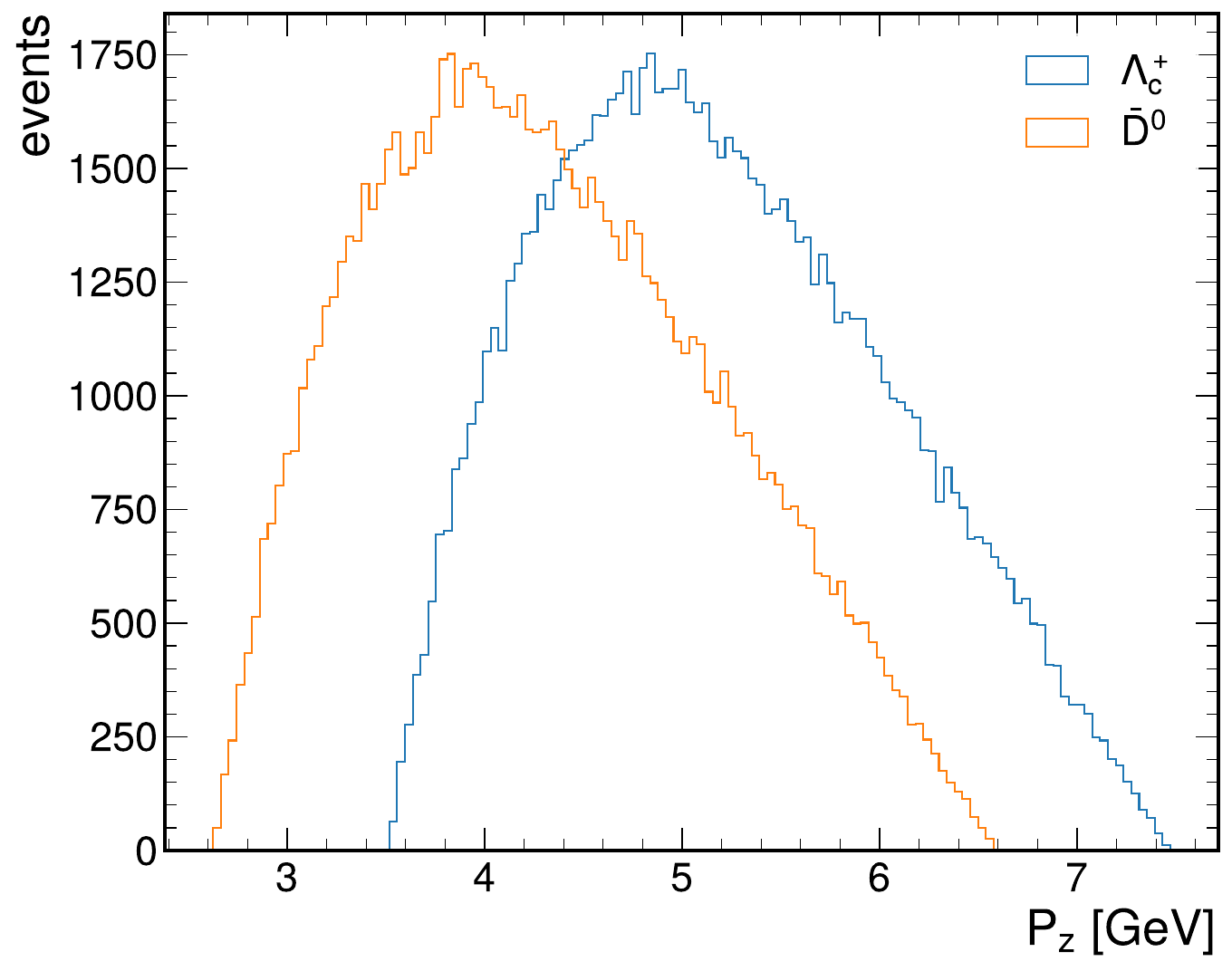}\includegraphics[width=0.5\textwidth]{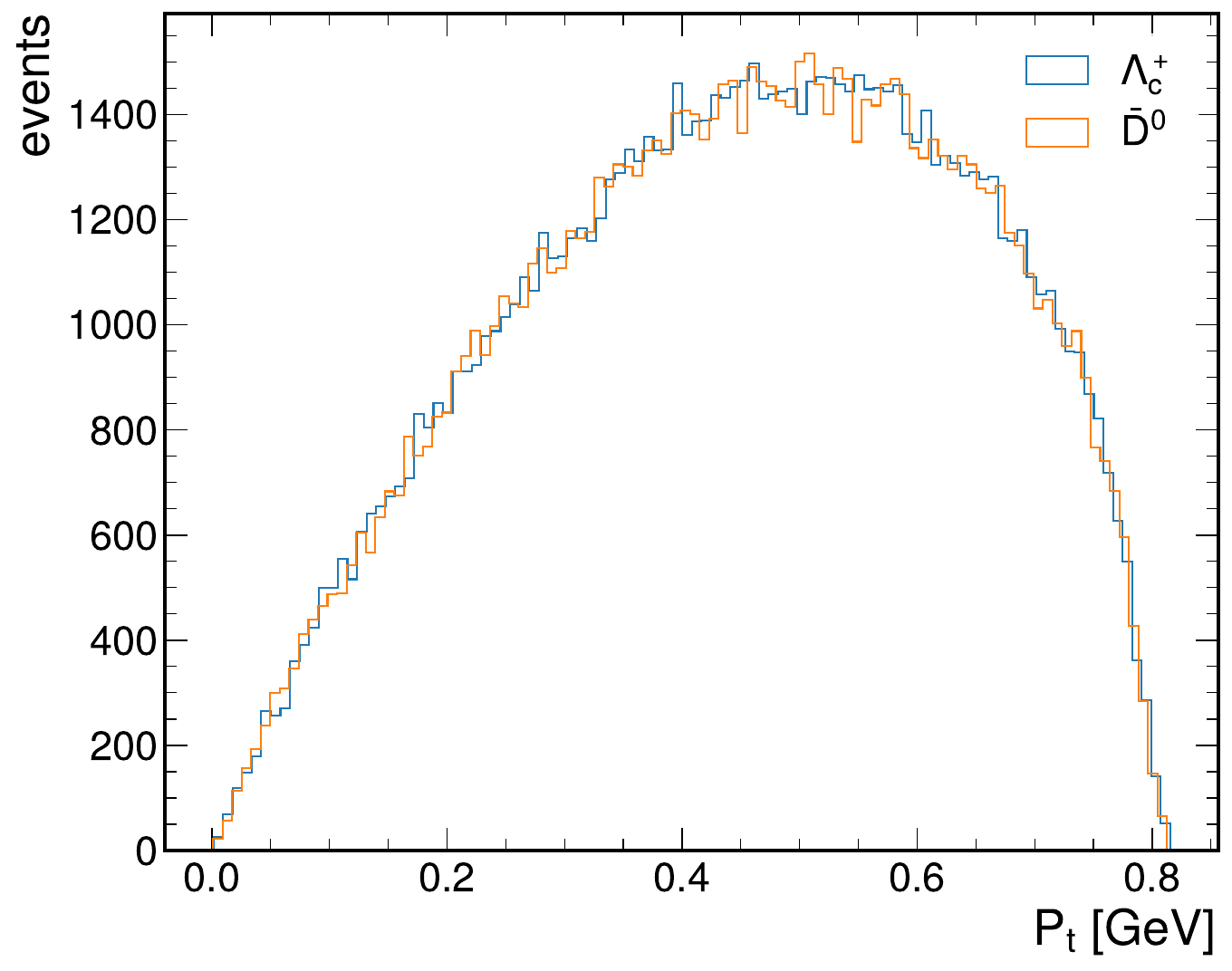}

     \includegraphics[width=0.5\textwidth]{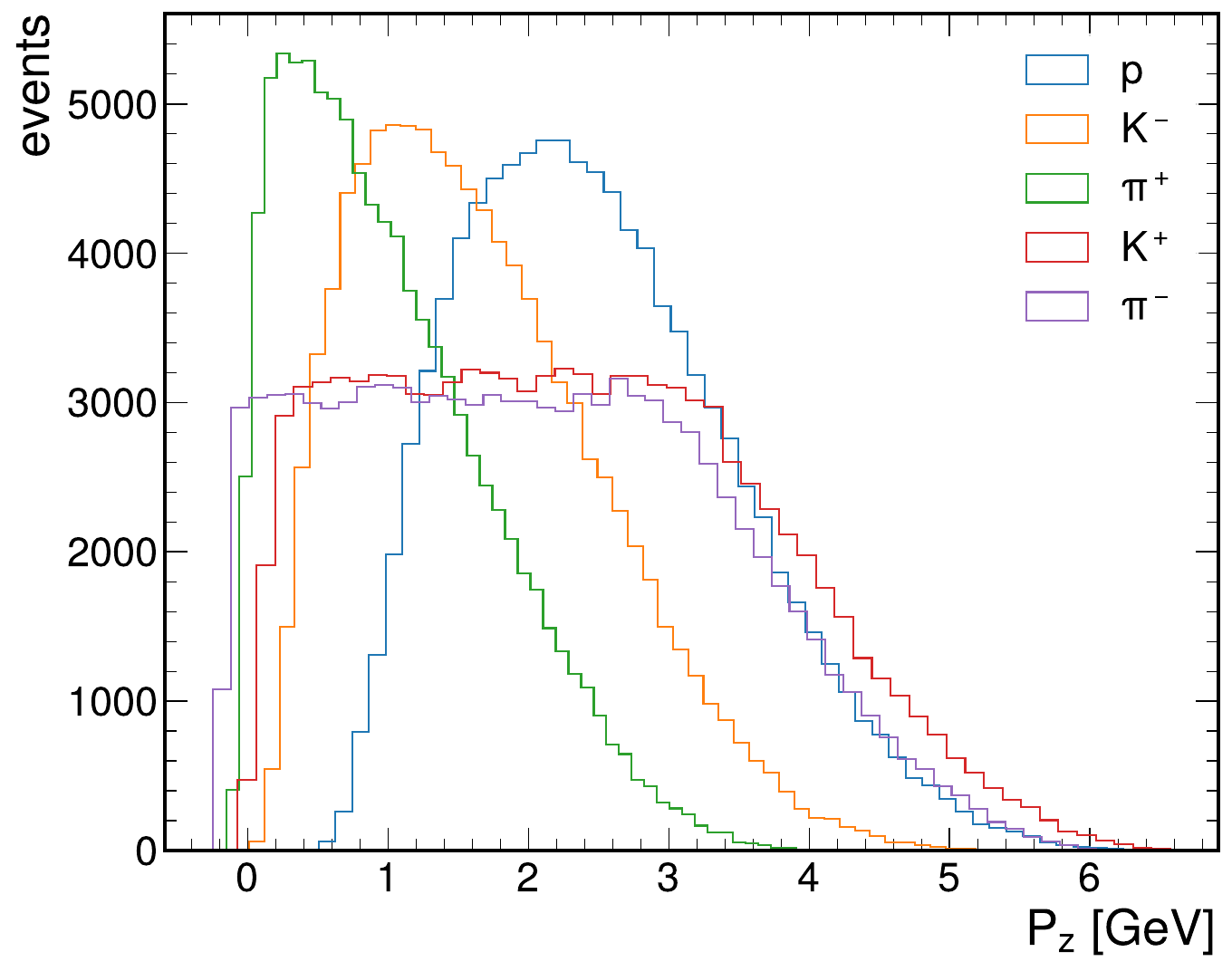}\includegraphics[width=0.5\textwidth]{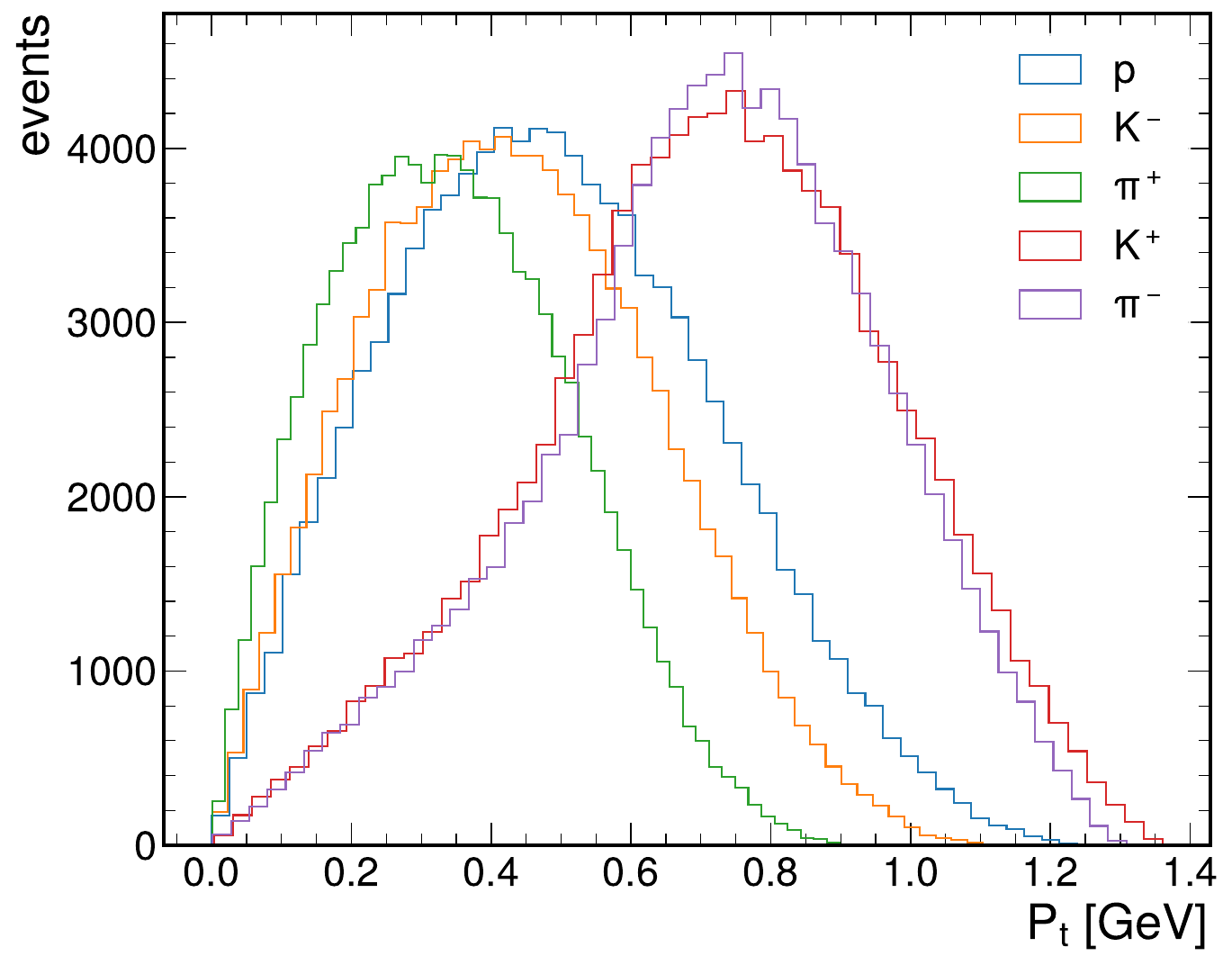}
     \caption{Top row: longitudinal (left) and transverse (right) momenta distributions for the generated charmed hadrons. Bottom row:  same, for their daughter particles}
     \label{fig:kinematics}
 \end{figure}

 To estimate the number of events in which either the $\Lambda_c^+$ or $\bar D^0$ can be measured through their decays, we first ran MC simulations of $ep\rightarrow e'\Lambda_c^+\bar D^0$ channel, in which the electron kinematic distributions correspond to Eq.~\ref{eq:epformula}.  The $\Lambda_c^+\bar D^0$ pairs were then produced with a uniform distribution in phase space given the combined four momentum of the virtual photon and proton.  We then allowed the simulated $\bar D^0$ to decay isotropically into a $K^+\pi^-$ pair, with the vertex-position given by an exponential distribution, taking into consideration the lifetime of the $\bar D^0$ and time dilation.  We did the same for the $\Lambda_c^+$, with the $K^-\pi^+p$ being generated with uniform probability within the 3-particle phase space.  The resulting momentum distributions for these simulated events are shown in Fig.~\ref{fig:kinematics}.
\subsection{Estimation of efficiency and acceptance}
Assuming that there is a combined detection efficiency and acceptance of around 80\% for each of the hadrons in the forward detector, and a similar efficiency for the electrons in the FT, approximately $(80\%)^3\approx52\%$ ($(80\%)^4\approx41\%$) of the events with $e'K^+\pi^-$ ($e'K^-\pi^+p$) in their final state will have all of these particles reconstructed.  We will further refine these estimates of the event-reconstruction  with using CLAS12's detector-response simulator package, GEMC.  
 % We then ran these events through CLAS12's event generator, GEMC, using default settings and found that only about 0.10\% of generated events had a reconstructed scattered electron in the FT and a reconstructed $K^-\pi^+p$ combination with invariant mass consistent with a $\Lambda_c^+$ decay.  For the $\bar D^0$ channel, only 0.52\% of generated events had a reconstructed scattered electron in the FT and a reconstructed $K^+\pi^-$ pair consistent with a $\bar D^0$ decay.  These values include both acceptance effects and detector efficiency.

 %\subsection{Branching ratios}
 %The branching ratios of 
 %   $\Lambda_c^+\rightarrow K^-\pi^+p$ and $\bar D^0\rightarrow K^+\pi^-$ listed in the PDG are respectively %$6.28\pm0.32\%$ and $=3.88\pm0.05\%$. 

 \subsection{Geometric acceptance of the MVTX}
\begin{figure}
    \centering
    \includegraphics[width=0.6\textwidth]{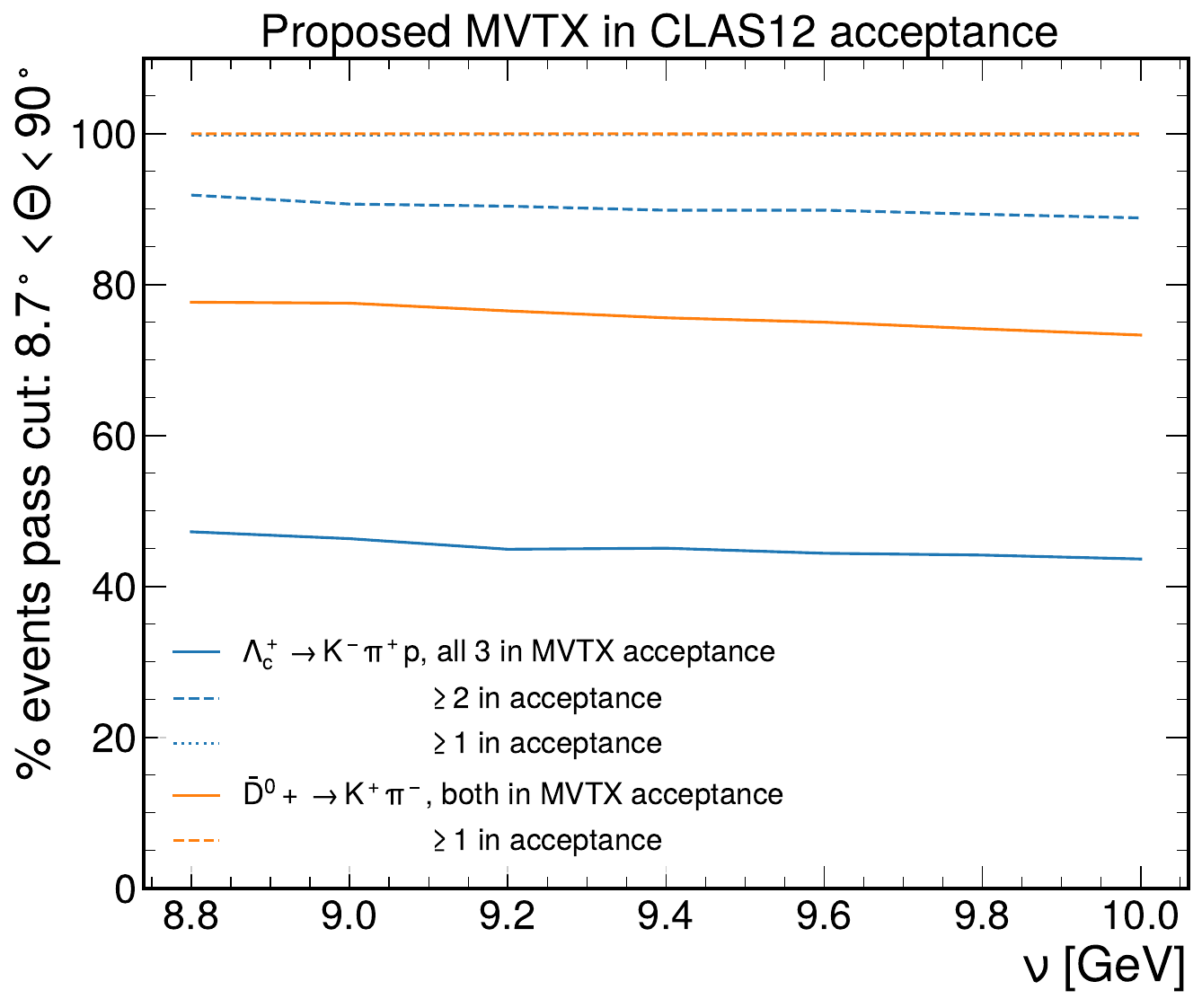}
    \caption{Overall geometric acceptance of the MVTX for detecting at least one of the daughter particles (dashed curves), at least two daughter particles (dotted curves) and all of the daughter particles (solid curves) for $\Lambda^+_c$ (blue) and $\bar D^0$ (orange).}
    \label{fig:MVTX_acceptance}
\end{figure}

To estimate the acceptance of the MVTX, we used the generated event sample described above.  In principle, determining the decay position of a charmed hadron requires at least one of its decay products to be within the proposed, nominal acceptance of MVTX at CLAS12: $8.7^\circ<\theta<90^\circ$.  However, as an alternative to reduce backgrounds from poorly reconstructed single tracks, it is possible to require at least two of the decay products to be reconstructed in the MVTX, or in the case of the $\Lambda_c^+$ decay, all three of the daughter particle candidates.  We show the overall geometric acceptance of the MVTX (that is the fraction of the generated events which a requisite number of the decay products are in the MVTX's geometric acceptance) of each of these scenarios in Fig.~\ref{fig:MVTX_acceptance}.  Almost all of the events have at least one daughter particle in the MVTX's acceptance.  Even in the most strict requirements (requiring all three daughter particles from the $\Lambda^+_c$ in the MVTX acceptance) would retain over 40\% of the events.  
 
\subsection{MVTX efficiency}
We estimate an 80\% efficiency for reconstructing each daughter particle in the MVTX, leading to $(80\%)^3\approx51\%$ efficiency for the $K^-\pi^+p$ decay and $(80\%)^3\approx64\%$ for the $K^+\pi^-$ channel, for events where all of the daughter particles are in the geometric acceptance of the MVTX.  A more precise estimate of this efficiency can be determined through future studies with Geant simulations.  

\subsection{Event retention after displaced-vertex cut}
To estimate how the thickness of the target would affect the efficiency of a displaced-vertex cut, we used the event-generator simulations described above.  The resolution for the vertex position of a track can depend on several factors including the magnetic field, the polar angle of the track, and the transverse momentum of the particle.  This will need to determined through further studies with Geant simulations.  In this exercise, we choose to use a conservative cut on $v_z$ at 250~$\mu$m from the downstream face of the target foil.  The efficiency of this cut can be calculated as:
\begin{equation}
    \epsilon_h(v_{z,\rm cut})=\frac{N(v^h_z>v_{z,\rm cut})}{N_{\rm tot}}
\end{equation}
where $N(v_z>v_{z,\rm cut})$ is the number of events where the reconstructed decay position of one of the charmed hadrons, $v^h_z$, (where the $h$ index represents either $\Lambda_c^+$ or $\bar D^0$), is greater than the cutoff, $v_{z,\rm cut}$.  The $v^h_z$ for each event is given by 
\begin{equation}
    v^h_z= v^{h,\rm gen}_z+v^e_z
\end{equation}
where $v^{h,\rm gen}_z$ is the $z$ position of the charmed-hadron decay from the event generator ($i.e.$, relative to the electron vertex), and $v^e_z$ is the electron vertex, which assigned uniformly to each event between $-t$ and $0$, where $t$ is the thickness of the foil.  

\begin{figure}[h!]
    \centering
    \includegraphics[width=0.6\textwidth]{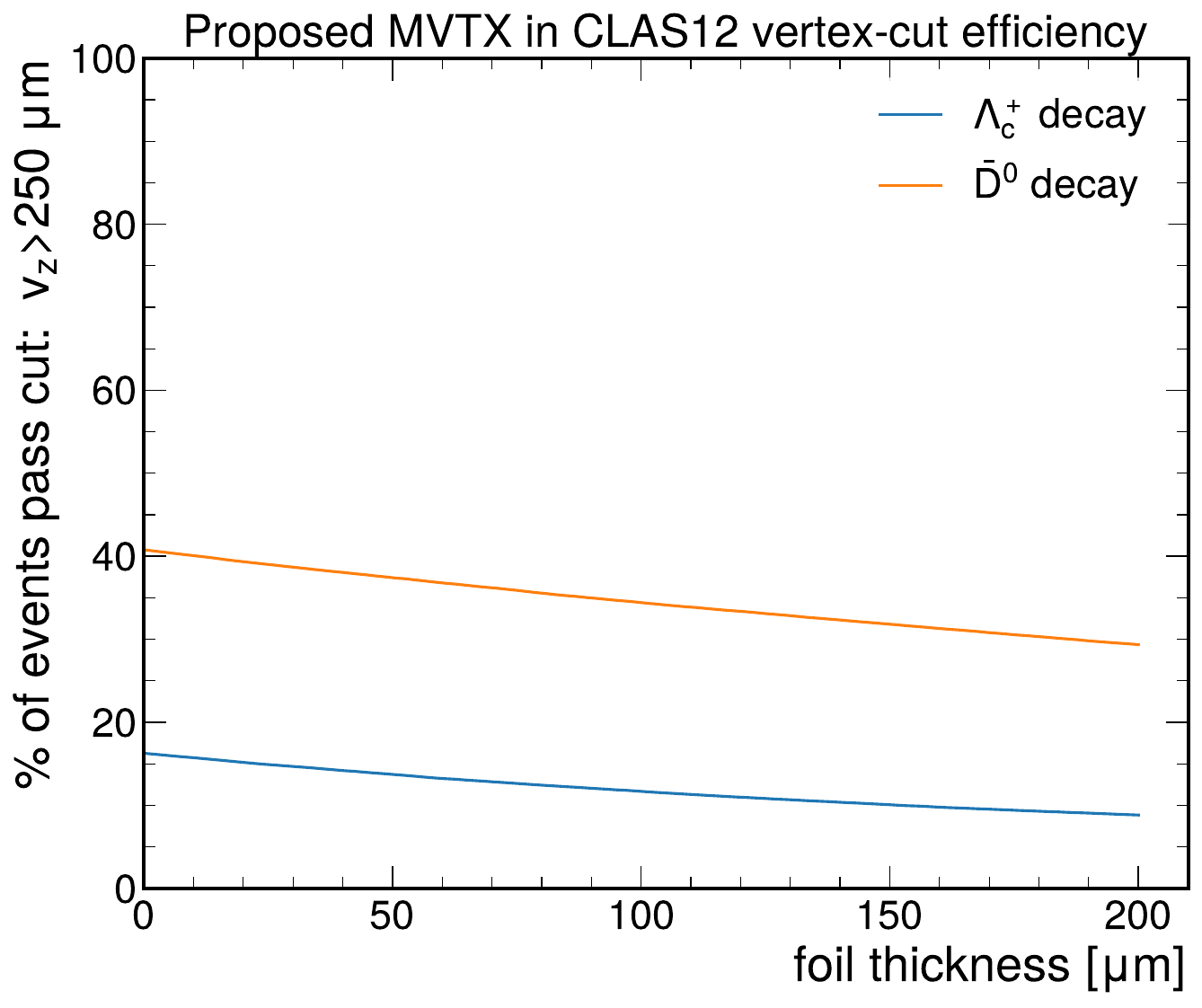}
    \caption{Efficiency for a displaced-vertex cut at 250~$\mu$m for the $\Lambda_c^+$ (blue) and $\bar D_0$ (orange) channels respectively.  For reference, the expected distance these particles travel is 130 (260) $\mu$m for $\Lambda_c^+$ ($\bar D^0$) at the mode energy, 4.9 (3.9) GeV.}
    \label{fig:vertex_cut_efficiency}
\end{figure}

The resulting cut efficiency, as a function of the foil thickness, is shown in Fig.~\ref{fig:vertex_cut_efficiency}. With 100 $\mu$m foils, about 12\% of $\Lambda_c^+$s and 34\% of $\bar D^0$s decay after the cutoff.  For reference, the CLAS12 run-group E experiment (on-going at the time of writing this letter of intent) uses a Pb target that is 143~$\mu$m thick. 

We intend to follow up these estimates with more refined tracking studies using realistic field maps provided by the CLAS12 magnet and Kalman-Filter reconstruction.

% \subsection{Background estimates from RGA}
% For an estimate of the number of background events, we looked for $K^+\pi^-$ and $K^-\pi^+p$ events in the RGA run period, in which the invariant mass of the decay-product candidates was within 3$\sigma$ of the expected mass of the charmed hadron, where $\sigma$ is the resolution in the invariant mass determined from the MC simulations: 7 MeV for $K^-\pi^+p$ events and 10 MeV for $K^+\pi^-$ events.  
% We found 381k events for the $K^+\pi^-$ channel and 216k $K^-\pi^+p$ events within 3$\sigma$ of their respective charmed hadron's masses.  If the $v_z$ distribution has non-Gaussian tails, it is difficult to know \textit{a priori} the reduction factor in the background due to the $v_z$ cut, but conservatively a 5~$\sigma$ cut in $v_z$ would be expected to reduce the background by at least 4 orders of magnitude.  

\subsection{Expected sample size and reach}
To estimate the number of events that can be measured for a given sample, we multiplied the total number of expected events by the branching ratio for a given channel, times the CLAS12 overall efficiency, the geometric acceptance of the MVTX, and the event retention rate of the displaced vertex cut, as detailed in the previous subsections.  %This yields, as a conservative estimate, 1.7k events with a reconstructed $\Lambda_c^+$ candidate and 11k events with a $\bar D^0$ candidate.  
This yields, as a conservative estimate, 340 events with a reconstructed $\Lambda_c^+$ candidate and 1.6k events with a $\bar D^0$ candidate.  
The factors that go into these calculations are summarized in Table~\ref{tab:calc}.

\begin{table}[]
    \centering
    \begin{tabular}{c|c|c|p{0.5\textwidth}}
      Quantity & $K^-\pi^+p$ & $K^+\pi^-$ & notes\\
      \hline
      free-proton cross-section & \multicolumn{2}{c|}{0.0082 nb} &  $\sigma_{\gamma p}$ calculated from Ref.~\cite{Du:2020bqj}, $\eta=0.5$, converted to $\sigma_{ep}$ (Eq.~\ref{eq:epformula})\\
      $Z/N$ & \multicolumn{2}{c|}{0.44} &  evaluated for $^{9}$Be \\
   Tot.~luminosity & \multicolumn{2}{c|}{1.3$\times10^8$ nb$^{-1}$} & 100 days, 50\% PAC efficiency, 10$^{35}$ cm$^{-2}$s$^{-1}$, achieved with 5-foil, 100 $\mu$m Be target at 85~nA; evaluated per nucleon\\
      Branching ratio &
       6.26\%  & 3.88\% & from PDG \\
      %CLAS12 efficiency & 0.10\% & 0.52\% & GEMC simulations\\
      CLAS12 efficiency & 41\% & 51\% & assuming 80\% per detected particle\\
      MVTX acceptance & 45\% & 76\% & all daughter particles in MVTX \\
      MVTX efficiency & 51\% & 64\% & assume 80\% reconstruction efficiency per daughter particle \\
      $v_z$ cut & 12\% & 34\% & assuming 100 $\mu$m per foil, 250 $\mu$m cut\\
      
      \hline
      %Tot. expected events & 1.7k & 11k \\
      Tot. expected signal events & 340& 1600&
    \end{tabular}
    \caption{Factors and assumptions that go into the calculations for the total expected yields (bottom row).}
    \label{tab:calc}
\end{table}

%We show in Fig.~\ref{fig:reach} the expected statistical precision of this experiment, under these assumptions, with 14 equally-spaced bins in $\nu$ (100 MeV each).  To calculate the statistical uncertainty in The uncertainty in each bin, we first estimated the number of events in each bin, $N_i$, which is given by
%\begin{equation}
 %   N_i=N_{\rm tot}\frac{\sigma_i\epsilon_i}{\sum\limits_{j\in\rm bins}\sigma_j\epsilon_j},
%\end{equation}
%where $N_{\rm tot}$ is the estimated total number of expected events in the combined $K^-\pi^+p$ and $K^+\pi^-$ channels passing all cuts, $\sigma_i$ is the $ep$ cross section in the $i^{\rm th}$ bin, and $\epsilon_i$ is the efficiency of detecting an electron in the FT\footnote{The FT has lower efficiency at electron energies just above the minimum detectable electron energy, and this is reflected in the lower statistical precision at large $\nu$}.  The estimated statistical precision on the measured cross section in each bin is given by the calculated cross section divided by $\sqrt{N_i}$.  

We have not studied background rates, as it heavily relies upon the background-rejection power of the vertex-position cuts, which requires detailed tracking studies with Geant simulations. These are outside the scope of this Letter of Intent.

%\begin{figure}
%    \centering
%    \includegraphics[width=0.7\textwidth]{Figures/reach.pdf}
%    \caption{Expected statistical precision of the experiment, compared to the calculations from Du \textit{et al.}~\cite{Du:2020bqj}, with $\eta=0.5$.}
%    \label{fig:reach}
%\end{figure}
%\input{Kinematics.tex}

%\input{Sketchs}

\FloatBarrier
\section{Summary and Conclusions}

The inclusion of measurements on open-charm production at JLab would complement the existing \jpsi~program, offering valuable insights into reaction mechanisms. Direct measurements of channels such as \opencharm~and \opencharmprime~would contribute to constraining the \jpsi~reaction mechanisms, shedding light on whether the open-charm coupled channel mechanism~\cite{Du:2020bqj} plays a significant role in the observed \jpsi~cross-sections. This constraint is crucial for a nuanced interpretation of \jpsi~data, challenging the commonly used vector-meson dominant model employed for extracting gravitational form factors of the nucleon.

To execute this program, the conclusion of operations for sPHENIX MVTX post-RHIC shutdown offers a promising opportunity. Using its capability to offer clear D-meson tagging via displaced vertex signatures, we suggest incorporating the sPHENIX MVTX geometry into the CLAS12 tracker. This integration, dependent on removing the standard vacuum scattering chamber and cryotarget, would involve installing a smaller vacuum chamber to accommodate a small solid target that would consist of multiple foils each $\approx$100$\mu$m thick.

The use of light solid targets like beryllium or carbon, with appropriate modeling, holds the potential to constrain corresponding cross-sections on a proton target. Additionally, an inclusive D-meson production program with light and heavy nuclear targets such as lead or uranium-oxide could provide insights into cold-nuclear matter effects with charm quarks, which could complement ongoing light-quark studies at JLab~\cite{CLAS:2021jhm,CLAS:2022asf}. All of these could complement and pave the way for proposed charm studies at the Electron-Ion Collider~\cite{Accardi:2012qut} and JLab at 22 GeV~\cite{Accardi:2023chb}.

We intend to follow up this LOI with detailed Geant detector simulations, encompassing signals and backgrounds, as well as thorough technical assessments of operating conditions in CLAS12, or other detector at JLab.

\bibliography{biblio.bib} 

\end{document}